\documentclass[12pt]{article}
\usepackage{latexsym}
\usepackage{amsmath}
\usepackage{amssymb}
\catcode`\@=11
\newcount\hour
\newcount\minute
\newtoks\amorpm \hour=\time\divide\hour by 60\minute
=\time{\multiply\hour by 60 \global\advance\minute by-\hour}
\edef\standardtime{{\ifnum\hour<12 \global\amorpm={am}%
        \else\global\amorpm={pm}\advance\hour by-12 \fi
        \ifnum\hour=0 \hour=12 \fi
        \number\hour:\ifnum\minute<10
        0\fi\number\minute\the\amorpm}}
\edef\militarytime{\number\hour:\ifnum\minute<10
0\fi\number\minute}
\def\draftlabel#1{{\@bsphack\if@filesw {\let\thepage\relax
   \xdef\@gtempa{\write\@auxout{\string
      \newlabel{#1}{{\@currentlabel}{\thepage}}}}}\@gtempa
   \if@nobreak \ifvmode\nobreak\fi\fi\fi\@esphack}
        \gdef\@eqnlabel{#1}}
\def\@eqnlabel{}
\def\@vacuum{}
\def\marginnote#1{}
\def\draftmarginnote#1{\marginpar{\raggedright\scriptsize\tt#1}}
\def\draft{
        \pagestyle{plain}
        \overfullrule=2pt
        \oddsidemargin -.1truein
        \def\@oddhead{\sl \phantom{\today\quad\militarytime} \hfil
        \smash{\Large\sl DRAFT} \hfil \today\quad\militarytime}
        \let\@evenhead\@oddhead
        \let\label=\draftlabel
        \let\marginnote=\draftmarginnote
        \def\ps@empty{\let\@mkboth\@gobbletwo
        \def\@oddfoot{\hfil \smash{\Large\sl DRAFT} \hfil}
        \let\@evenfoot\@oddhead}
        \def\@eqnnum{(\theequation)\rlap{\kern\marginparsep\tt\@eqnlabel}%
        \global\let\@eqnlabel\@vacuum}  }

\renewcommand{\thefootnote}{\fnsymbol{footnote}}

\def\appendix#1{\addtocounter{section}{1}\setcounter{equation}{0}
\renewcommand{\thesection}{\Alph{section}}
\section*{Appendix \thesection\protect\indent \parbox[t]{11.15cm}{#1}}
\addcontentsline{toc}{section}{Appendix \thesection\ \ \ #1}}

\def \bi{\bibitem}
\def \la {\label}

\def \b {\beta}

\def\be{\begin{equation}}
\def\ee{\end{equation}}

\catcode`\@=11

\def\bea{\begin{eqnarray}}
\def\eea{\end{eqnarray}}
\def\beann{\begin{eqnarray*}}
\def\eeann{\end{eqnarray*}}
\def\beq{\begin{equation}}
\def\eeq{\end{equation}}
\def\ba{\begin{array}}
\def\ea{\end{array}}
\def\ben{\begin{enumerate}}
\def\een{\end{enumerate}}

 \def \la {\label}
 \def\be{\begin{equation}}
\def\ee{\end{equation}}

\def \la {\label}

\font\mybb=msbm10 at 11pt

\def\bb#1{\hbox{\mybb#1}}

\def\bR {\bb{R}}

\def\e  {\epsilon}

\def \ee {\epsilon}

\def \bi{\bibitem}
\def\a{\alpha }

\def \b {\beta}

\def\be{\begin{equation}}
\def\ee{\end{equation}}

\def \bi {\bibitem}
\def \la{\label}

\begin{document}
\date{November 2002}
\begin{titlepage}
\begin{center}
\hfill UB-ECM-PF-07-17 \\

\vspace{3.0cm} {\Large \bf Supersymmetric heterotic string
backgrounds }
\\[.2cm]

\vspace{1.5cm}
 {\large  U.~Gran$^1$,
 G.~Papadopoulos$^2$, and D.~Roest$^3$}

\vspace{0.5cm}

${}^1$ Fundamental Physics\\
Chalmers University of Technology\\
SE-412 96 G\"oteborg, Sweden\\

\vspace{0.5cm}
${}^2$ Department of Mathematics\\
King's College London\\
Strand,
London WC2R 2LS, UK\\

\vspace{0.5cm}
${}^3$ Departament Estructura i Constituents de la Materia \\
    Facultat de F\'{i}sica, Universitat de Barcelona \\
    Diagonal 647, 08028 Barcelona, Spain \\

\end{center}

\vskip 1.5 cm
\begin{abstract}

 We present the main features of the solution of the gravitino and dilatino Killing spinor
equations derived in hep-th/0510176 and hep-th/0703143 which have led to  the classification of
 geometric types of all type I  backgrounds. We then apply
these results to the supersymmetric backgrounds of the heterotic string.
In particular, we solve the gaugino Killing spinor equation  together with the other two Killing spinor equations of the theory.
We also use our results to classify all supersymmetry conditions of ten-dimensional gauge theory.

\end{abstract}
\end{titlepage}
\newpage
\setcounter{page}{1}
\renewcommand{\thefootnote}{\arabic{footnote}}
\setcounter{footnote}{0}

\setcounter{tocdepth}{0}

\setcounter{section}{0}
\setcounter{subsection}{0}

%
%

The effective theory of the heterotic string can be described by a type I supergravity which includes higher curvature corrections that
are organized in an $\a'$ and $g_s$ expansion. The $\a'$ corrections can be computed by a sigma model loop calculation
and modify the field equations of the theory. In addition, the anomaly cancelation mechanism modifies the
Bianchi identity of the three-form field strength. Up to and including  two-loops in the sigma model computation,
 the Killing spinor equations of the effective theory \cite{strominger, hullb, eric} can be written as
\bea
 {\cal D} \e = \hat\nabla\e+{\cal O}(\a'^2)=0~,~~~{\cal A} \e = (d\Phi-{1\over2} H)\e+{\cal O}(\a'^2)=0~,~~~ {\cal F} \e  = F \e+{\cal O}(\a'^2)=0~,
\la{kse}
\eea
where we have suppressed the spacetime indices and the gamma matrix dependence\footnote{
We use the notation of \cite{het} where a more detailed description can be found.}.
The first equation is the gravitino Killing spinor equation for a metric connection $\hat\nabla$
with torsion the three-form field strength $H$. The second equation is the dilatino Killing spinor equation, where $\Phi$
is the dilaton, and the last is the gaugino Killing spinor equation, where $F$ is the gauge field strength.
It is clear that these Killing spinor equations are as those expected from type I supergravity to the
order indicated.
The only difference is
that $H$ is not closed. In particular  it is modified at one-loop due to the Green-Schwarz anomaly cancellation mechanism as
\bea
dH=- \tfrac14 \a' \big( {\rm tr} \check R^2- {\rm tr} F^2\big)+{\cal O}(\a'^2)~,
\la{anb}
\eea
where $\check R$ is the curvature of the connection $\check \nabla=\nabla-{1\over2}H$ with torsion $-H$.
The field equations of the effective theory up to and including two-loops in the sigma model computation can been found
in \cite{hull}.

We have presented the Killing spinor equations of the effective theory as an $\a'$-expansion
 containing high order curvature terms. These are not known   to all orders. However, the Killing spinor
equations and the modified Bianchi identity have also been viewed as exact, see e.g.~\cite{li}. In such a case, all ${\cal O}(\a'^2)$ terms
are neglected and the remaining $\a'$ dependence suppressed. The two-loop contribution
to the Einstein equation and in particular the $\check R^2$ term is needed for consistency \cite{gillard}, see also \cite{lust}.
There is also a growing literature on geometries with skew-symmetric torsion in differential geometry, see e.g.
\cite{howegpc}-\cite{stefanspin7}.

In \cite{het, typeI}, applying the spinorial method of \cite{ggp}, the gravitino and dilatino Killing spinor equations (\ref{kse})
were solved in all cases and
the underlying geometry of the spacetime was presented. The gaugino Killing spinor equation was not solved explicitly. This is
because in most cases it does not affect the geometry of spacetime. It is easy to see that the gaugino Killing spinor equation
is ``decoupled'' from the gravitino and dilatino ones. However, the gauge field strength $F$ contributes in the modified Bianchi
identity \eqref{anb} for $H$. Since  in the solution of the gravitino
and dilatino Killing spinor equations in
 \cite{het, typeI} it was {\it not} assumed that $dH=0$, it was not deemed necessary to solve the gaugino Killing spinor equation.
 One exception to this  are  the backgrounds with $\hat R=0$ for which one needs $dH=0$ to argue that the spacetime  is a group manifold.
But if $F\not=0$,  then the modified Bianchi identity  may give $d H\not=0$
and so the geometry may be deformed away from  that of a group manifold. Nevertheless the ten-dimensional spacetime is parallelizable
with respect to a metric connection with skew-symmetric torsion. It can be shown that such manifolds are either Lorentzian Lie groups or a product
of the Lorentzian Lie group with $S^7$ \cite{jfofhet}.

To incorporate the gaugino Killing spinor equation with the other two Killing spinor equations and to
consider the effect that the anomaly cancelation mechanism has on the geometry, we shall solve the
gaugino Killing spinor equation in all cases. As a consequence, we solve the supersymmetry
condition $F\e=0$ for all gauge theories up to ten dimensions. We find more cases than those that have
appeared in the literature so far, see e.g.
 \cite{corrigan}.

Before we proceed to do this let us recall the essential ingredients  of the classification of
geometric types of supersymmetric backgrounds in
\cite{het, typeI}. The first step is the observation that the integrability condition
of the gravitino Killing spinor equation gives
\bea
\hat R\e=0
\eea
which implies that either the isotropy group, ${\rm Stab}(\e_1,\dots, \e_L)$,  of the parallel spinors,
$\e_1,\dots, e_L$, is a proper subgroup of the
holonomy
group\footnote{Note that the $\a'$ corrections do not change the holonomy of the
supercovariant connection up to and including two-loops in the sigma model
expansion.} $Spin(9,1)$, or the isotropy group is $\{1\}$ and $\hat R=0$. In the latter case the backgrounds are parallelizable,
and if $dH=0$ they are group manifolds. The complete list of isotropy groups of spinors in $Spin(9,1)$
has been given in \cite{typeI}, for previous  work see \cite{josea}. This list characterizes all the solutions of the gravitino Killing spinor equation.

Suppose we have a solution of the gravitino Killing spinor equation with $L$ parallel spinors.  To
 solve the dilatino Killing spinor equation,
 consider  the
group
\bea
\Sigma({\cal P})={\rm Stab}({\cal P})/{\rm Stab}(\e_1,\dots, \e_L)~,
\eea
where ${\cal P}$ is the $L$-plane spanned by all parallel spinors. One can then  show that
the first Killing spinor $\zeta_1$, which is  a linear combination of  $\e_1,\dots, \e_L$, can be chosen
as a representative of the orbits of $\Sigma({\cal P})$ on ${\cal P}$.
Having found the first Killing spinor, one can see that the second $\zeta_2$ can be chosen  as the representatives of the orbits of
${\rm Stab}({\cal P}_1)\subset \Sigma({\cal P})$ on ${\cal P}/{\cal P}_1$, where ${\cal P}_1=\bR<\zeta_1>$. This can be repeated to
find representatives
for all the spinors. For $N\geq L/2$, it more convenient to find the representatives of the normals
rather than the Killing spinor themselves. Using the machinery described and applying the spinorial geometry technique
  the dilatino Killing spinor equation
can be  solved  in {\it all} cases.
The list of all $\Sigma({\cal P})$ groups has been given in
\cite{typeI}, see also  table 1.

\begin{table}[ht]
 \begin{center}
\begin{tabular}{|c|c|c|}\hline
${\mathrm {Parallel~spinors}}$ & ${\mathrm Stab}(\e_1,\dots,\e_L)$ & $\Sigma({\cal P})$
 \\ \hline \hline
$1+e_{1234}$ & $Spin(7)\ltimes\bR^8$& $Spin(1,1)$ \\
\hline
$1$ &  $SU(4)\ltimes\bR^8$&$Spin(1,1)\times U(1)$
\\ \hline
$1, i(e_{12}+e_{34})$ & $Sp(2)\ltimes\bR^8$&$Spin(1,1)\times SU(2)$
\\ \hline
$1, e_{12}$ & $(SU(2)\times SU(2))\ltimes\bR^8$&$Spin(1,1)\times Sp(1)\times Sp(1)$
\\ \hline
$1, e_{12}, e_{13}+e_{24}$ & $SU(2)\ltimes\bR^8$&$Spin(1,1)\times Sp(2)$
\\ \hline
$1, e_{12}, e_{13}$ & $U(1)\ltimes\bR^8$&$Spin(1,1)\times SU(4)$
\\ \hline
$1, e_{12}, e_{13}, e_{14}$ & $\bR^8$&$Spin(1,1)\times Spin(8)$
\\ \hline \hline
$1+e_{1234}, e_{15}+e_{2345}$ & $G_2$&$Spin(2,1)$
\\ \hline
$1, e_{15}$ & $SU(3)$&$Spin(3,1)\times U(1)$
\\ \hline
$1, e_{15}, e_{12}, e_{25}$ & $SU(2)$&$Spin(5,1)\times SU(2)$
\\ \hline
$1, e_{ij}, e_{i5},~ i,j=1,\dots,4$ & $\{1\}$&$Spin(9,1)$
\\ \hline
\end{tabular}
\end{center}
\caption{
In the  columns are a basis in the space of parallel spinors, their isotropy groups in $Spin(9,1)$ and the $\Sigma({\cal P})$ groups, respectively.
The basis of parallel spinors is  always real. So if a complex spinor is given as a basis spinor  it is understood that one should
always take the real and imaginary parts. }
\end{table}

To solve all the Killing spinor equations (\ref{kse}), it is convenient to consider these equations in
the following order
\bea
{\rm gravitino}  \rightarrow {\rm gaugino} \rightarrow {\rm dilatino}
\nonumber
\eea
Starting from a solution
of the gravitino Killing spinor equation, we determine those parallel spinors which also solve the
gaugino Killing spinor equation. Finally, given a solution of the gravitino and gaugino Killing spinor equations,
we shall describe how all solutions of the dilatino Killing spinor equation can be found.
There may be backgrounds for which the dilatino Killing spinor equation has more solutions than the gravitino one.
However, since we are interested in the solution of all Killing spinor equation, the order that we have chosen
to solve them is not essential. In addition the gravitino Killing spinor equation has a direct topological and geometric significance
and so it makes sense to consider it first.

Starting from  a solution of the gravitino Killing spinor equation, one can argue that the spinors that solve
the gaugino Killing spinor equation can be selected up to $\Sigma({\cal P})$ transformations, where ${\cal P}$ is the space
of parallel spinors as before. The argument
for this is similar to that presented in \cite{typeI} for selecting the spinors that solve the dilatino Killing spinor equation.
 An additional simplifying factor here is
 that $F\e=0$ has a non-trivial solution iff the spinors that
solve the gaugino Killing spinor equation have a non-trivial isotropy group in $Spin(9,1)$. This is because $F$ is a Lie algebra
valued two-form, and the space of two forms at every spacetime point   can be identified with the Lie algebra
$\mathfrak{spin}(9,1)$. Thus $F\e=0$ can be viewed as an invariance condition for $\e$ under $\mathfrak{spin}(9,1)$ rotations
 generated by $F$. It turns out that
this imposes strong restrictions on the solutions.
In particular, it follows that the solutions of both gravitino and gaugino Killing spinor equations
can be expressed in terms of  bases as those given in table 1 for the parallel spinors.
Let ${\cal P}_F$ denote the plane that spans the solution of the gravitino and gaugino Killing spinor equations in each case.
It turns out that in all cases $\Sigma({\cal P}_F)$, which  we  use to select the solutions of the dilatino Killing spinor equation,
 can always be identified with a $\Sigma({\cal P})$ group of table 1. Thus the results of \cite{typeI}
 can then be used to solve the dilatino Killing spinor equation in all cases.

 To describe the results, let
 $N_F$  be the number solutions of the gravitino
and gaugino Killing spinor equations.  The number $N$ of Killing spinors, i.e.~the number of solutions
to all Killing spinor equations, is $N\leq N_F\leq L$. In what follows, we shall describe the solution of the gravitino Killing spinor
equation by the isotropy of the parallel spinors in $Spin(9,1)$. Then, we shall give all gaugino Killing spinor equations
and their solutions\footnote{The solutions of the gaugino Killing spinor equations can always be described by saying $F\in \mathfrak{h}$ for some Lie
subalgebra $\mathfrak{h}\subset \mathfrak{so}(9,1)$. This is a short-hand notation to indicate that $F$ takes values in the subbundle
of the bundle of two-forms $\Lambda^2(M)$ of the spacetime defined by the adjoint representation of $\mathfrak{h}$.
 For example $F\in \mathfrak{spin}(7)\oplus \bR^8$ means that there are gauge Lie algebra
valued one- and two-forms $\a$ and $\b$, respectively such that $F=e^-\wedge \a+\b$ and
$\b_{ij}={1\over2}\phi_{ij}{}^{kl} \b_{kl}$, where $\phi$ is the  fundamental $Spin(7)$ four-form.}. In addition,
given a solution of both gravitino and gaugino Killing spinor equations, we shall
indicate which dilatino Killing spinor equations remain to be solved. In all cases, the solution of the
latter is given in \cite{typeI}.

\vskip 0.2cm
\underline{{$Spin(7)\ltimes\bR^8$}}

\vskip 0.2cm
There is only one possibility because $N_F=L=1$.
{\small
\begin{itemize}
\item $N_F=L=1$, ${\cal F}(1+e_{1234})=0\Longrightarrow F\in \mathfrak{spin}(7)\oplus\bR^8$
\begin{itemize}
\item $N=1$:~~ ${\cal A}(1+e_{1234})=0$
\end{itemize}
\end{itemize}}
\vskip 0.2cm
\underline{{$SU(4)\ltimes\bR^8$}}
\vskip 0.2cm
There are two possibilities which arise with $N_F=1$ and $N_F=L=2$. $N_F=1$ is
the same as in the previous case. So the only new case is
{\small
\begin{itemize}
\item $N_F=L=2$, ${\cal F}1=0\Longrightarrow F\in \mathfrak{su}(4)\oplus\bR^8$, $\Sigma({\cal P}_F)=Spin(1,1)\times U(1)$
\begin{itemize}
\item $N=1$:~~ ${\cal A}(1+e_{1234})=0$
\item $N=2$:~~  ${\cal A}1=0$
\end{itemize}
\end{itemize}}

The $\Sigma({\cal P}_F)=Spin(1,1)\times U(1)$ group can be read from table 1.

\vskip 0.2cm
\underline{{$Sp(2)\ltimes\bR^8$}}
\vskip 0.2cm
Three possibilities   arise with $N_F=1,2,3$. The $N_F=1,2$ are the same as those described in the previous case. So the only new case  is
{\small
\begin{itemize}
\item $N_F=L=3$, ${\cal F}1={\cal F}(e_{12}+e_{34})=0\Longrightarrow F\in \mathfrak{sp}(2)\oplus\bR^8$, $\Sigma({\cal P}_F)=Spin(1,1)\times SU(2)$
\begin{itemize}
\item $N=1$:~~ ${\cal A}(1+e_{1234})=0$
\item $N=2$:~~  ${\cal A}1=0$
\item $N=3$:~~  ${\cal A}1={\cal A}(e_{12}+e_{34})=0$
\end{itemize}
\end{itemize}}

\vskip 0.2cm
\underline{{$(SU(2)\times SU(2))\ltimes\bR^8$}}
\vskip 0.2cm
Four possibilities arise with $N_F=1,2,3,4$. The first three $N_F=1,2,3$ are as in the previous case. So the only new case is
{\small
\begin{itemize}
\item $N_F=L=4$, ${\cal F}1={\cal F}e_{12}=0\Longrightarrow F\in \mathfrak{su}(2)\oplus\mathfrak{su}(2) \oplus\bR^8$, $\Sigma({\cal P}_F)=Spin(1,1)\times Sp(1)\times Sp(1)$
\begin{itemize}
\item $N=1$:~~ ${\cal A}(1+e_{1234})=0$
\item $N=2$:~~  ${\cal A}1=0$
\item $N=3$:~~  ${\cal A}1={\cal A}(e_{12}+e_{34})=0$
\item $N=4$:~~  ${\cal A}1={\cal A}e_{12}=0$
\end{itemize}
\end{itemize}}

\vskip 0.2cm
\underline{{$SU(2)\ltimes\bR^8$}}
\vskip 0.2cm

Five possibilities arise with $N_F=1,2,3,4,5$. The first four $N_F=1,2,3,4$ are as in the previous case. So, the only new case is
{\small
\begin{itemize}
\item $N_F=L=5$, ${\cal F}1={\cal F}e_{12}= {\cal F}(e_{13}+e_{24})=0\Longrightarrow F\in \mathfrak{su}(2) \oplus\bR^8$, $\Sigma({\cal P}_F)=Spin(1,1)\times Sp(2)$
\begin{itemize}
\item $N=1$:~~ ${\cal A}(1+e_{1234})=0$
\item $N=2$:~~  ${\cal A}1=0$
\item $N=3$:~~  ${\cal A}1={\cal A}(e_{12}+e_{34})=0$
\item $N=4$:~~  ${\cal A}1={\cal A}e_{12}=0$
\item $N=5$:~~  ${\cal A}1={\cal A}e_{12}={\cal A}(e_{13}+e_{24})=0$
\end{itemize}
\end{itemize}}

\vskip 0.2cm
\underline{{$U(1)\ltimes\bR^8$}}
\vskip 0.2cm

Six possibilities arise with $N_F=1,2,3,4,5, 6$. The first five $N_F=1,2,3,4, 5$ are as in the previous case. So, the only new case is
{\small
\begin{itemize}
\item $N_F=L=6$, ${\cal F}1={\cal F}e_{12}= {\cal F}e_{13}=0\Longrightarrow F\in \mathfrak{u}(1) \oplus\bR^8$, $\Sigma({\cal P}_F)=Spin(1,1)\times SU(4)$
\begin{itemize}
\item $N=1$:~~ ${\cal A}(1+e_{1234})=0$
\item $N=2$:~~  ${\cal A}1=0$
\item $N=3$:~~  ${\cal A}1={\cal A}(e_{12}+e_{34})=0$
\item $N=4$:~~  ${\cal A}1={\cal A}e_{12}=0$
\item $N=5$:~~  ${\cal A}1={\cal A}e_{12}={\cal A}(e_{13}+e_{24})=0$
\item $N=6$:~~  ${\cal A}1={\cal A}e_{12}={\cal A}e_{13}=0$
\end{itemize}
\end{itemize}}

\vskip 0.2cm
\underline{{$\bR^8$}}
\vskip 0.2cm
Seven possibilities arise with $N_F=1,2,3,4,5, 6, 8$. The first six $N_F=1,2,3,4, 5, 6$ are as in the previous case. The $N_F=7$
does not occur. In fact if $N_F=7$, then the gaugino Killing spinor equation admits an additional solution which gives $N_F=8$. So, the only new case is
{\small
\begin{itemize}
\item $N_F=L=8$, ${\cal F}1={\cal F}e_{12}= {\cal F}e_{13}={\cal F} e_{14}=0\Longrightarrow F\in \bR^8$, $\Sigma({\cal P}_F)=Spin(1,1)\times Spin(8)$
\begin{itemize}
\item $N=1$:~~ ${\cal A}(1+e_{1234})=0$
\item $N=2$:~~  ${\cal A}1=0$
\item $N=3$:~~  ${\cal A}1={\cal A}(e_{12}+e_{34})=0$
\item $N=4$:~~  ${\cal A}1={\cal A}e_{12}=0$
\item $N=5$:~~  ${\cal A}1={\cal A}e_{12}={\cal A}(e_{13}+e_{24})=0$
\item $N=6$:~~  ${\cal A}1={\cal A}e_{12}={\cal A}e_{13}=0$
\item $N=7$:~~ ${\cal A}1={\cal A}e_{12}={\cal A}e_{13}={\cal A}(e_{14}-e_{23})=0$
\item $N=8$:~~ ${\cal A}1={\cal A}e_{12}={\cal A}e_{13}={\cal A}e_{14}=0$
\end{itemize}
\end{itemize}}
\noindent This completes the description of cases that arise from parallel spinors that have non-compact isotropy groups. For compact isotropy groups one has

\vskip 0.2cm
\underline{{$G_2$}}
\vskip 0.2cm
There are two possibilities, $N_F=1$ and $N_F=L=2$.
{\small
\begin{itemize}
\item $N_F=1$, ${\cal F}(1+e_{1234})=0\Longrightarrow F\in \mathfrak{spin}(7)\oplus\bR^8$
\begin{itemize}
\item $N=1$:~~ ${\cal A}(1+e_{1234})=0$
\end{itemize}
\item $N_F=L=2$, ${\cal F}(1+e_{1234})={\cal F}(e_{15}+e_{2345})=0\Longrightarrow F\in \mathfrak{g}_2$, $\Sigma({\cal P}_F)=Spin(2,1)$
\begin{itemize}
\item $N=1$:~~ ${\cal A}(1+e_{1234})=0$
\item $N=2$:~~  ${\cal A}(1+e_{1234})={\cal A}(e_{15}+e_{2345})=0$
\end{itemize}
\end{itemize}}

\vskip 0.2cm
\underline{{$SU(3)$}}
\vskip 0.2cm
In this case $N_F=1, 2$ and $N_F=L=4$. The $N_F=3$ case does not occur because the gaugino Killing spinor equations has an additional solution
giving $N_F=L=4$. There are two possibilities that occur for $N_F=2$.
{\small
\begin{itemize}
\item $N_F=1$, ${\cal F}(1+e_{1234})=0\Longrightarrow F\in \mathfrak{spin}(7)\oplus\bR^8$
\begin{itemize}
\item $N=1$:~~ ${\cal A}(1+e_{1234})=0$
\end{itemize}
\item $N_F=2$, ${\cal F}(1+e_{1234})={\cal F}(e_{15}+e_{2345})=0\Longrightarrow F\in \mathfrak{g}_2$, $\Sigma({\cal P}_F)=Spin(2,1)$
\begin{itemize}
\item $N=1$:~~ ${\cal A}(1+e_{1234})=0$
\item $N=2$:~~  ${\cal A}(1+e_{1234})={\cal A}(e_{15}+e_{2345})=0$
\end{itemize}
\item $N_F=2$, ${\cal F}1=0\Longrightarrow F\in \mathfrak{su}(4)\oplus\bR^8$, $\Sigma({\cal P}_F)=Spin(1,1)\times U(1)$
\begin{itemize}
\item $N=1$:~~ ${\cal A}(1+e_{1234})=0$
\item $N=2$:~~  ${\cal A}1=0$
\end{itemize}
\item $N_F=L=4$, ${\cal F}1={\cal F}e_{15}=0\Longrightarrow F\in \mathfrak{su}(3)$, $\Sigma({\cal P}_F)=Spin(3,1)\times U(1)$
\begin{itemize}
\item $N=1$:~~ ${\cal A}(1+e_{1234})=0$
\item $N=2$:~~  ${\cal A}1=0$
\item $N=2$:~~  ${\cal A}(1+e_{1234})={\cal A}(e_{15}+e_{2345})=0$
\item $N=3$:~~ ${\cal A} 1={\cal A}(e_{15}+e_{2345})=0$
\item $N=4$:~~  ${\cal A} 1={\cal A} e_{15}=0$
\end{itemize}
\end{itemize}}

\vskip 0.2cm
\underline{{$SU(2)$}}
\vskip 0.2cm
In this case $N_F=1, 2, 3, 4$ and $N_F=L=8$. The $N_F=5,6,7$ cases do not occur because the gaugino Killing spinor equation has
 additional solutions
giving $N_F=L=8$. The range of $N_F$ is expected but it is not a trivial result. To show this one has to substitute the
spinors that occur in the dilatonic Killing spinor equation for $SU(2)$ parallel spinors in \cite{typeI}
to the gaugino Killing spinor equation and eliminate  several cases. For example for $N_F=4$, there are six possible choices for
solutions of the gaugino Killing spinor equation but in fact only two of these give exactly four solutions. The rest
restrict $F$ to take values in $\mathfrak{su}(2)$ and so the dilatino Killing spinor equation admits eight solutions.
There are two possibilities that occur for $N_F=2,4$.
{\small
\begin{itemize}
\item $N_F=1$, ${\cal F}(1+e_{1234})=0\Longrightarrow F\in \mathfrak{spin}(7)\oplus\bR^8$
\begin{itemize}
\item $N=1$:~~ ${\cal A}(1+e_{1234})=0$
\end{itemize}
\item $N_F=2$, ${\cal F}(1+e_{1234})={\cal F}(e_{15}+e_{2345})=0\Longrightarrow F\in \mathfrak{g}_2$, $\Sigma({\cal P}_F)=Spin(2,1)$
\begin{itemize}
\item $N=1$:~~ ${\cal A}(1+e_{1234})=0$
\item $N=2$:~~  ${\cal A}(1+e_{1234})={\cal A}(e_{15}+e_{2345})=0$
\end{itemize}
\item $N_F=2$, ${\cal F}1=0\Longrightarrow F\in \mathfrak{su}(4)\oplus\bR^8$, $\Sigma({\cal P}_F)=Spin(1,1)\times U(1)$
\begin{itemize}
\item $N=1$:~~ ${\cal A}(1+e_{1234})=0$
\item $N=2$:~~  ${\cal A}1=0$
\end{itemize}
\item $N_F=3$, ${\cal F}1={\cal F}(e_{12}+e_{34})=0\Longrightarrow F\in \mathfrak{sp}(2)\oplus\bR^8$, $\Sigma({\cal P}_F)=Spin(1,1)\times SU(2)$
\begin{itemize}
\item $N=1$:~~ ${\cal A}(1+e_{1234})=0$
\item $N=2$:~~  ${\cal A}1=0$
\item $N=3$:~~  ${\cal A}1={\cal A}(e_{12}+e_{34})=0$
\end{itemize}
\item $N_F=4$, ${\cal F}1={\cal F}e_{15}=0\Longrightarrow F\in \mathfrak{su}(3)$, $\Sigma({\cal P}_F)=Spin(3,1)\times U(1)$
\begin{itemize}
\item $N=1$:~~ ${\cal A}(1+e_{1234})=0$
\item $N=2$:~~  ${\cal A}1=0$
\item $N=2$:~~  ${\cal A}(1+e_{1234})={\cal A}(e_{15}+e_{2345})=0$
\item $N=3$:~~ ${\cal A} 1={\cal A}(e_{15}+e_{2345})=0$
\item $N=4$:~~  ${\cal A} 1={\cal A} e_{15}=0$
\end{itemize}
\item $N_F=4$, ${\cal F}1={\cal F}e_{12}=0\Longrightarrow F\in \mathfrak{su}(2)\oplus\mathfrak{su}(2)\oplus\bR^8 $, $\Sigma({\cal P}_F)=Spin(1,1)\times Sp(1)\times Sp(1)$
\begin{itemize}
\item $N=1$:~~ ${\cal A}(1+e_{1234})=0$
\item $N=2$:~~  ${\cal A}1=0$
\item $N=3$:~~  ${\cal A}1={\cal A}(e_{12}+e_{34})=0$
\item $N=4$:~~  ${\cal A}1={\cal A}e_{12}=0$
\end{itemize}
\item $N_F=L=8$, ${\cal F}1={\cal F}e_{12}={\cal F}e_{15}={\cal F}e_{25}=0\Longrightarrow F\in \mathfrak{su}(2)$, $\Sigma({\cal P}_F)=Spin(5,1)\times  SU(2)$

The possibilities that arise for the gaugino Killing spinor equation have been given in \cite{typeI}. There are solutions
for $1\leq N\leq 8$. The $N=7$ does not occur provided that $dH=0$.
\end{itemize}}

\vskip 0.2cm
\underline{{$\{1\}$}}
\vskip 0.2cm
In this case all spinors are parallel. Since $\Sigma({\cal P})=Spin(9,1)$, the spinors that are solutions
of the gaugino Killing spinor equation are precisely those that appear as parallel in table 1. As a consequence
 the $\Sigma({\cal P}_F)$
groups  coincide with the $\Sigma({\cal P})$ groups in each case. In particular we have that
  $N_F=1,2,3,4,5,6,8,16$. The cases $N_F=7$ and $8<N_F<16$ do not occur. There are two possibilities
with $N_F=2,4,8$.  Therefore, the condition  $F\e=0$ requires that $F$ as a spacetime  two-form
takes values in the subalgebras
\bea
&&\mathfrak{spin}(7)\oplus \bR^8 \; (1)~,~~~\mathfrak{su}(4)\oplus \bR^8 \; (2)~,~~~ \mathfrak{sp}(2)\oplus \bR^8  \; (3)~, ~~~
\mathfrak{su}(2)\oplus \mathfrak{su}(2)\oplus \bR^8  \; (4)~,~~~
\cr
&&\mathfrak{su}(2)\oplus \bR^8 \; (5)~, ~~~\mathfrak{u}(1)\oplus \bR^8 \;  (6)~, ~~~\bR^8 \; (8)~,~~~
\cr
&&\mathfrak{g}_2 \; (2)~,~~~\mathfrak{su}(3) \;  (4)~,~~~\mathfrak{su}(2) \; (8)~,~~~  \{0\} \; (16)~.
\eea
of $\mathfrak{spin}(9,1)$, where the number in parenthesis denotes $N_F$. These are {\it all} the
supersymmetry conditions of ten-dimensional gauge theory on parallelizable space and in particular on $\bR^{9,1}$.
There are more conditions than the BPS conditions given in \cite{corrigan}.
Moreover the dilatino Killing spinor equations are those described in \cite{typeI} for each group
$\Sigma({\cal P}_F)=\Sigma({\cal P})$ listed in table 1.
In the description of cases with compact holonomy groups above, there is
a degree of repetition. This is done   for clarity.

It is clear that after solving the gravitino and gaugino Killing spinor equations, the possibilities that arise
for the solutions of the dilatino Killing spinor equations depend on the group $\Sigma({\cal P}_F)$. Since as we have mentioned
all the $\Sigma({\cal P}_F)$ groups coincide with some   $\Sigma({\cal P})$, the solutions of the dilatino Killing spinor
equation can be read
off from those of \cite{typeI}. In table 2, we summarize all the cases that arise emphasizing the multiplicity
of possibilities for a given number  $N$ of Killing spinors. This multiplicity is defined as the
number of different dilatino KSE that one solves for a given $N$. Each case typically leads to a different spacetime
geometry.

\begin{table}[ht]
 \begin{center}
\begin{tabular}{|c|c|}\hline
   $\Sigma({\cal P}_F)$ &$N$
 \\ \hline \hline
  $Spin(1,1)$& 1$^1$ \\
\hline $Spin(1,1)\times U(1)$&1$^1$, 2$^1$
\\ \hline
$Spin(1,1)\times SU(2)$&1$^1$, 2$^1$, 3$^1$
\\ \hline
$Spin(1,1)\times Sp(1)\times Sp(1)$&1$^1$, 2$^1$, 3$^1$, 4$^1$
\\ \hline
$Spin(1,1)\times Sp(2)$&1$^1$, 2$^1$, 3$^1$, 4$^1$, 5$^1$
\\ \hline
$Spin(1,1)\times SU(4)$&1$^1$, 2$^1$, 3$^1$, 4$^1$, 5$^1$, 6$^1$
\\ \hline
$Spin(1,1)\times Spin(8)$&1$^1$, 2$^1$, 3$^1$, 4$^1$, 5$^1$, 6$^1$, 7$^1$, 8$^1$
\\ \hline \hline
$Spin(2,1)$& 1$^1$, 2$^1$
\\ \hline
$Spin(3,1)\times U(1)$&1$^1$, 2$^2$, 3$^2$, 4$^1$
\\ \hline
$Spin(5,1)\times SU(2)$&1$^1$, 2$^2$, 3$^3$, 4$^6$, 5$^3$, 6$^2$, 7$^1$, 8$^1$
\\ \hline
$Spin(9,1)$& 8$^2$, 10$^1$, 12$^1$, 14$^1$, 16$^1$
\\ \hline
\end{tabular}
\end{center}
\caption{ In the  columns are the $\Sigma({\cal P}_F)$ groups that
arise from the solution of the gravitino and gaugino Killing spinor
equations and the number $N$  of supersymmetries, respectively. The number in superscript indicates the different cases that arise
 in the dilatino Killing spinor equation for a given $N$.
}
\end{table}

It is easily observed that if $\Sigma({\cal P })$ is associated with a non-compact stability subgroup, then there is
a unique dilatino Killing spinor equation that arises for a given $N$. However, if the stability group is compact,
then several cases can arise. The most involved is that of $\Sigma({\cal P}_F)=Spin(5,1)\times SU(2)$ for  $N=4$ for
 which there are six different
types of dilatino Killing spinor equations that arise up to gauge transformations of the Killing spinor equations.

We have presented the list of geometric conditions on the spacetime  that  arise from the solution of the Killing spinor equations
in all cases. It is not apparent that there will always be examples of spacetimes that satisfy all these conditions. The existence
of solutions is another problem which has to be examined separately\footnote{ This is a similar to the
situation that arises in the  Berger classification  between the list of holonomies of
 irreducible simply connected Riemannian manifolds and the proof that manifolds with such holonomy actually exist.}.
We have already seen in \cite{typeI} that imposing $dH=0$ and the field equations together with mild assumptions on the holonomy of $\hat\nabla$ imply that large classes of possibilities do not occur. To extend this
to the case that the Bianchi identity is modified,  we expand the dilaton as
\bea
\Phi=\sum^\infty_{n=0} (\a')^n \Phi_n~,
\eea
and similarly for the other fields.
It is clear that $(g_0, \Phi_0, H_0, A_0)$ must satisfy the Killing spinor equations and field equations
with $dH_0=0$. So these are subject to the conditions mentioned above and large classes of descendants  do not exist.
It remains to see how the $\a'$ corrections modify the result.  A calculation of this type has been
done for holonomy $SU(2)$ \cite{callan, howegpb}, holonomy $SU(3)$   \cite{gillard} and holonomy $\{1\}$ \cite{jfofhet} backgrounds. The result
of the first computation is tuned to a particular example. For the $SU(3)$ case, the computation is more general but still the
spacetime is restricted  to be a  metric product $M=\bR^{3,1}\times B$.   In the holonomy $\{1\}$ case, the gravitational contribution
to the anomaly vanishes and it can be shown that
the gauge field contribution does not deform the spacetime away from a group manifold.

We have presented a complete description of the solutions of gravitino, gaugino and dilatino Killing spinor equations
and we have found the geometry of all supersymmetric backgrounds of type I supergravity.
It is clear that the next step is to find examples of solutions in all cases. Many are known already. For the fundamental string \cite{gibbons}
and pp-wave propagating in $\bR^8$ solutions,
 ${\rm hol}(\hat\nabla)=\bR^8$ and all parallel  spinors are Killing. For the NS5-brane solution,
${\rm hol}(\hat\nabla)=SU(2)$ and again all parallel Killing spinors are Killing, and similarly for the heterotic
5-brane \cite{callan}. The holonomy of 5-brane intersections can be found in \cite{tesch}, see also \cite{gauntlett, lustb}. For the background in \cite{volkov} which has applications
in gauge theory \cite{mn}, ${\rm hol}(\hat\nabla)=SU(3)$
\cite{tseytlin} and  all parallel spinors are Killing,   and similarly for the Calabi-Yau compactifications.
The understanding of the geometric properties   of all solutions allows us to investigate them  beyond a case by case basis.
The Bianchi identities, field equations and additional assumptions on the holonomy put strong restrictions on the existence
of solutions.
For ${\rm Stab}(\e_1, \dots, \e_L)$  non-compact, if one assumes ${\rm hol}(\hat\nabla)={\rm Stab}(\e_1,\dots, \e_L)$,
 $dH=0$ and the field equations, then
the gravitino Killing spinor equation implies the dilatino one and all parallel spinors are Killing, so
there are no descendants \cite{typeI}.
We can also require that the transverse spaces to the light-cone directions in the non-compact isotropy group case
or the base space of the principal fibration in the compact isotropy group case to be compact.
For example, this is desirable in the context of compactifications with fluxes.
Such an assumption imposes restrictions on the geometry of spacetime. In particular, it has been shown that under  certain conditions that
smooth  backgrounds of the type $\bR^{9-2n,1}\times B^{2n}$ and ${\rm hol}({\hat \nabla})\subset SU(n)$ exist, iff
$B^{2n}$ is a Calabi-Yau, i.e.~$H=0$ and the dilaton is constant \cite{ivanovgp}. This extends in a straightforward manner
to all backgrounds with ${\rm Stab}(\e_1,\dots,\e_L)$ non-compact.

Another direction to extend  these investigations is to the type II common sector backgrounds. Since the Killing spinor equations
in this case are two copies of the type I, it is clear that we have solved all Killing spinor equations of one of the copies.
It would be interesting to find out what additional conditions one has to impose such that there are additional
Killing spinors in the other copy. This would extend the work of \cite{commonII}.

\vskip 0.5cm \noindent{\bf Acknowledgements} \vskip 0.1cm The work of U.G.~is funded by the Swedish Research Council.
D.R.~is supported by the European EC-RTN project
MRTN-CT-2004-005104, MCYT FPA 2004-04582-C02-01 and CIRIT GC
2005SGR-00564.

\end{document}